\documentclass[12pt]{article}
\usepackage{amsfonts,amsmath,amssymb,epsf}
\usepackage{graphicx,color}
\newcommand{\al}{\alpha'}
\newcommand{\de}{\partial}
\newcommand{\be}{\begin{equation}}
\newcommand{\ba}{\begin{eqnarray}}
\newcommand{\ea}{\end{eqnarray}}
\newcommand{\ee}{\end{equation}}

\newcommand{\f}{\frac}
\newcommand{\s}{\sqrt}
\newcommand{\vp}{\varphi}

\newcommand{\ti}{\tilde}

\newcommand{\ddd}{\cdot\cdot\cdot}
\newcommand{\no}{\nonumber \\}
\newcommand{\la}{\langle}
\newcommand{\lb}{\rangle}
\newcommand{\ep}{\epsilon}

\begin{document}

\begin{titlepage}
\thispagestyle{empty}



\begin{center}
\noindent{\Large \textbf{Holographic Calculation of Boundary Entropy}}\\
\vspace{2cm} \noindent{Tatsuo
Azeyanagi$^{a}$\footnote{e-mail:aze@gauge.scphys.kyoto-u.ac.jp},\!
Andreas Karch$^{b}$\footnote{e-mail:karch@phys.washington.edu},\!
Tadashi
Takayanagi$^{a}$\footnote{e-mail:takayana@gauge.scphys.kyoto-u.ac.jp},\!
and Ethan G. Thompson$^{b}$\footnote{e-mail:egthomps@u.washington.edu}}\\
\vspace{1cm}

 {\it $^{a}$Department of Physics, Kyoto University, Kyoto 606-8502, Japan \\
 $^{b}$Department of Physics, University of Washington, Seattle, Wa, 98195, USA}

\vskip 2em
\end{center}

\begin{abstract}
We use the holographic proposal for calculating entanglement
entropies to determine the boundary entropy of defects in strongly
coupled two-dimensional conformal field theories. We study several
examples including the Janus solution and show that the boundary entropy
extracted from the entanglement entropy as well as its more
conventional definition via the free energy agree with each other.
Maybe somewhat surprisingly we find that, unlike in the case of a
conformal field theory with boundary, the entanglement entropy for a
generic region in a theory with defect carries detailed information
about the microscopic details of the theory. We also argue that the
g-theorem for the boundary entropy is closely related to the strong
subadditivity of the entanglement entropy.

\end{abstract}

\end{titlepage}

\newpage

\section{Introduction}
\setcounter{equation}{0} Quantum systems with defects or boundaries
often show interesting physical behaviors. For example, impurities
in various materials have been one of the major subjects of research
in condensed matter physics. In such systems, we expect that the
whole system (both the bulk and the boundary) flows towards a fixed point of the renormalization
group in the low energy limit, where we can employ the powerful
method of conformal field theory. The boundary or
defect preserves a part of the conformal symmetry at the fixed
point.

One of the important quantities which characterizes properties of the
defects or boundaries is known as the boundary entropy (or, equivalently, the ground
state degeneracy $g$) \cite{Affleck:1991tk}. This measures the
degrees of freedom localized at a given defect and is a boundary
analogue of the central charge $c$. Recently, it has also been
pointed out that the boundary entropy can be regarded as the finite
part of the entanglement entropy \cite{Calabrese:2004eu}.

In general, the theory is strongly interacting at the RG fixed point and
sometimes it is very difficult to calculate physical quantities
like the boundary entropy. However, if the theory has a holographic
dual, we can compute many quantities rather simply by using the dual gravity
description. The most tractable examples will be the ones for which
we can apply the AdS/CFT correspondence.

The purpose of this paper is to holographically compute the boundary
entropy of 2d conformal field theories with defects using several
methods (for early discussions refer to \cite{Yamaguchi:2002pa}). The holographic calculation of entanglement entropy has been
recently formulated \cite{Ryu:2006bv,Ryu:2006ef}. This allows us to find the boundary 
entropy from the entanglement entropy, in
addition to using a probe computation of the boundary entropy at high
temperature.

The organization of this paper is as follows. In section
\ref{BEandEE}, we present a brief summary of the definition and
properties of the boundary entropy. We will also work out a close
relation between the g-theorem and the strong subadditivity of
entanglement entropy. In section \ref{holographicBE}, we perform the
holographic computations of the boundary entropy using both
probe configurations and fully back-reacted geometries (in particular for the Janus solution).
In section \ref{conclusion} we summarize our conclusions. In
appendix \ref{appendix2point}, we present the calculations of two
point functions and the boundary entropy of a free scalar field in the
presence of the interface.

\section{Boundary Entropy and Entanglement Entropy}
\label{BEandEE} \setcounter{equation}{0} In this paper, we are
interested in two dimensional conformal field theories (2d CFTs) in
the presence of a conformal defect. If we define the time and space
coordinate by $(t,x)$, then we can consider a time-like defect which
is situated at $x=0$. The defect is called conformal if a linear
combination of two Virasoro symmetries in the bulk is preserved. We
will refer to a CFT with such a defect as a defect conformal field
theory (DCFT) (e.g. see the review part of \cite{Bachas:2001vj}).
Generically, there are extra propagating degrees of freedom
localized on the defect. However, it is also possible to construct a
system with no new degrees of freedom on the defect. Such a theory
is called an interface CFT (ICFT). A simple example of an ICFT is a
compactified scalar field $\phi(t,x)$ whose radius jumps at the
defect. An interesting quantity which characterizes a system with a
conformal boundary or defect is the boundary entropy, $S_{bdy}$.
$S_{bdy}$ is related to the ground state degeneracy $g$
\cite{Affleck:1991tk} as we explain below.

\subsection{Boundary Entropy and the $g$-function}

Consider a 2d CFT with periodic Euclidean time, $t\sim
t+2\pi$. We assume $x$ is also compactified on a large circle with
radius $L\gg1$. When we introduce a defect at $x=0$, the
partition function of this system on a torus behaves as
  \be \lim_{L\to \infty} Z_{torus}=
e^{-LE_0+S_{bdy}}, \ee where $E_0$ is the ground state energy when
we regard $x$ as the Euclidean time direction. The quantity
$S_{bdy}$ is called the boundary entropy. This is motivated by the
observation that $S_{bdy}$ is the entropy when we artificially treat
$L$ as the temperature and $-\f{\log Z_{torus}}{L}$ as the free
energy. Originally, the boundary entropy was defined in a 2d system
with a conformal boundary in \cite{Affleck:1991tk}. However, the
DCFT can be equivalently described by two copies of the system with
boundary via the doubling trick, as discussed, for example, in
\cite{Bachas:2001vj} or in appendix A of this paper.

The quantity $g$, defined by \be g \equiv e^{S_{bdy}}, \ee represents the
ground state degeneracy. We can extend the idea of the boundary
entropy to non-conformal systems and define the $g$-function.
According to the $g$-theorem \cite{Affleck:1991tk}, the $g$-function
is a monotonically decreasing function with respect to the length
scale $l$,  \be \f{d}{dl}\log g(l)\leq
0, \label{gthfor} \ee in analogy to the $c$-function and $c$-theorem.

\subsection{Boundary Entropy from Entanglement Entropy}
\label{befromee}

Recently, it was found that the boundary entropy is actually related
to a physical entropy, the entanglement entropy
\cite{Calabrese:2004eu}. To define the entanglement entropy, we
first divide the system into two parts $A$ and $B$. Accordingly, the
total Hilbert space is factorized as $H=H_A\otimes H_B$. Next we
introduce the reduced density matrix $\rho_A=\mbox{Tr}_B\rho$ for
the subsystem  $A$  by tracing out $H_B$. Finally, the entanglement
entropy is defined as the von-Neumann entropy for $\rho_A$ \be
S_A=-\mbox{Tr}\rho_A\log \rho_A. \ee

Consider an infinitely long system and define the subsystem $A$ by
the finite interval with length $l$. The subsystem $B$ is defined to
be the complement of $A$. Then the entanglement entropy $S_A$ can be
computed to be \cite{Holzhey:1994we} \be S_A=\f{c}{3}\log \f{l}{a},
\label{bulk} \ee where $c$ is the central charge of the total system
and $a$ is the UV cut off (i.e. lattice spacing).

In the presence of a conformal boundary with boundary entropy
$\log g$, this is modified as follows \cite{Calabrese:2004eu}
 \be
S_A=\f{c}{6}\log \f{l}{a}+\log g . \label{entbg} \ee Because the
boundary cuts off half of the space, we have the coefficient
$\f{c}{6}$ instead of $\f{c}{3}$.

When we consider a conformal defect which is situated at the middle
of the interval $A$, we can regard the system as two copies of a
BCFT by the doubling trick. This leads to
the following result
 \be
S_A=\f{c}{3}\log \f{l}{a}+\log g . \label{entg} \ee  To see the
relation (\ref{entg}) quickly, let us remember that in the 2d CFT
$S_A$ can be found from the formula \be S_A=-\f{\de}{\de
n}\mbox{Tr}\rho_A^n\bigl|_{n=1} =-\f{\de}{\de
n}\left[\f{Z_n}{(Z_1)^n}\right]\Bigl|_{n=1}, \label{formulaz}\ee
where $Z_n$ is the partition function on the n-sheeted Riemann
surface with a cut along the interval $A$ \cite{Calabrese:2004eu}.
The important point is that both the original two dimensional space
and the $n$-sheeted one both have a single connected boundary. Thus
the ratio $\f{Z_n}{(Z_1)^n}$ is proportional to the factor
$g^{1-n}$, which leads to the formula (\ref{entg}).

On the other hand, if the defect is not located at the midpoint
of the interval, the entanglement entropy cannot be determined only
from $c$ and $g$, but rather it depends on the details of
the theory. This is because we cannot relate this DCFT setup to the
BCFT setup by the folding trick, as the quantity we are interested in
does not have the reflection symmetry about the defect. In other
words, it is not possible to find a conformal map from the
$n$-sheeted Riemann surface defined by $v^n=\f{w-l_1}{w+l_2}$ with
the defect at Re$~w=0$, to a single complex plane ${\bf C}$ with a
straight defect line except for $l_1=l_2$, which means that the defect
is at the midpoint of the interval.

\subsection{Strong Subadditivity and g-theorem}

It is intriguing to see if we can obtain useful properties of the
boundary entropy from the basic properties of entanglement entropy.
One of the most important inequalities satisfied by any entanglement
entropy is the strong subadditivity constraint (e.g. refer to the review
part of \cite{Casini:2004bw,Hirata:2006jx}). It is represented by
the inequality \be S_{A}+S_B \geq S_{A\cap B}+S_{A\cup B}.
\label{SS} \ee The holographic derivation of strong subadditivity
has been given in \cite{Headrick:2007km,Hirata:2006jx}.

\begin{figure}[htbp]
\begin{center}
  \hspace*{0.5cm}
  \includegraphics[height=3cm]{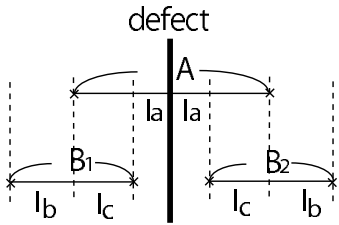}
  \caption{The simple setup of $A$ and $B$ at a common fixed time. Notice that
  both $A$ and $B$ live in the same one dimensional space.}\label{gth}
\end{center}
\end{figure}

It was shown in \cite{Casini:2004bw} that the entropic analogue of
the $c$-theorem follows from this relation. Therefore, it is natural
to ask if the $g$-theorem can also be derived from this condition.
Let us start with a simple setup (see fig.\ref{gth}.) in a defect
CFT. $A$ is defined by $[-l_a-l_c,\ l_a+l_c]$ and $B$ is defined by
the two intervals $[-l_a-l_b-l_c,-l_a]$ and $[l_a,\ l_a+l_b+l_c]$.
In this case, by substituting (\ref{entg}) into the strong
subadditivity constraint (\ref{SS}), we obtain \ba && \f{c}{3}\log
2(l_a+l_c) +\log g(2(l_a+l_c))+2\cdot \f{c}{3}\log (l_b+l_c) \no &&\
\ \ \  \geq 2\cdot \f{c}{3}\log l_c +
\f{c}{3}\log(2(l_a+l_b+l_c))+\log g(2(l_a+l_b+l_c)).\label{ineq} \ea
Then in the limit $l_b\to 0$ we find \be \f{d}{dl}\log
g(l)\bigl|_{l=2(l_a+l_c)}\leq
\f{c}{6}\left(\f{2}{l_c}-\f{1}{l_a+l_c}\right). \ee By taking the
limit $l_a\to 0$, we obtain the bound \be \f{d\log g(l)}{dl}\leq
\f{c}{3l}.\ee Even though this is not enough to prove the
$g$-theorem (\ref{gthfor}), we can at least say that the $g$-theorem
is non-trivially
consistent with the strong subadditivity.\\

\begin{figure}[htbp]
\begin{center}
  \hspace*{0.5cm}
  \includegraphics[height=3cm]{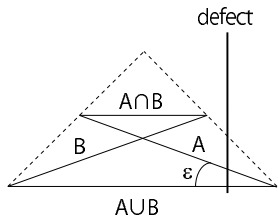}
  \caption{The relativistic setup of $A$ and $B$. The vertical and
  horizontal directions represent the time
  and space coordinates, respectively. The dotted light-like triangle
   delimits the causal future.}\label{gboost}
\end{center}
\end{figure}

To see the relation to the g-theorem more clearly, we need to cancel
the log terms in (\ref{ineq}). This can be done by considering a
relativistic setup as in fig.\ref{gboost}. Notice that by requiring
Lorentz invariance, the Hilbert space $H_A$ for $A$ depends only on
the causal future (or past) of $A$ and remains the same under any
deformation which preserves it. If the Lorentz invariant length of
$A$ is denoted by $|A|$, we can easily show $|A|\cdot |B|=|A\cap
B|\cdot |A\cup B|$. Owing to this relation, the authors in
\cite{Casini:2004bw} were able to prove the $c$-theorem from the
strong subadditivity condition for this setup. Indeed, strong subadditivity
leads to \be S(l_1)+S(l_2)\geq S(l_3)+S(l_4), \ee where we assume
$l_1l_2=l_3l_4$ and $l_4\leq l_1, l_2\leq l_3$. This is equivalent
to the concavity of the entropy as a function of $\log l$ i.e. \be
l\f{d}{dl}\left(l\f{dS(l)}{dl}\right)\leq 0.\label{entc}\ee Noting
that $c(l)=3l\f{dS(l)}{dl}$, it is clear from (\ref{bulk}) that the inequality
(\ref{entc}) is precisely the entropic $c$-theorem.

Now we return to the relation to the $g$-theorem and thus we assume
that the bulk region is conformal. If we again employ the choice of
subsystems and the defect line as described in fig.\ref{gboost}, the
bulk log terms are completely canceled out. Since the other part of
the entanglement entropy can be regarded as an entropic g-function,
we simply obtain \be \log g (A)\geq \log g (A\cup B). \label{ddd}
\ee This indeed agrees with the g-theorem in a particular case. In
this way, we have learned that strong subadditivity for the
entanglement entropy is closely related to the $g$-theorem for
DCFTs. We leave a further study of this issue as a future problem.


\section{Holographic Boundary Entropy of Defect}
\label{holographicBE} \setcounter{equation}{0} An interesting class
of 2d CFTs, in which the boundary entropy can be calculated using
its relation to the entanglement entropy, is those 2d CFTs which
have a dual description in terms of a higher dimensional
gravitational theory on an asymptotically AdS$_3$ background. A
method to calculate entanglement entropies in theories with
gravitational duals has been found in \cite{Ryu:2006bv,Ryu:2006ef}.
This equates the entanglement entropy of a spatial region ${\cal M}$
in the field theory bounded by $\partial {\cal M}$ with $\frac{A}{4
G_N^{(3)}}$, where $A$ is the area of a minimal area surface ending on $\partial {\cal M}$ and in a constant time slice of the 3d bulk  . $G_N^{(3)}$
is the 3d Newton constant. We will apply this formula here to
known examples of DCFTs with a gravitational dual.

In particular, we will look at three different systems. The first system
we study is a Randall-Sundrum (RS) like toy model
\cite{Randall:1999vf,Karch:2000ct} of a brane coupled to gravity,
which for a certain range of tensions has a dual description in
terms of a DCFT. Since this model, in its simplest form, has not
been embedded in string theory or any other consitent
theory of quantum gravity, we don't know precisely what the
DCFT is (and whether it exists at all).
But the advantage is that in this case we can calculate
both the entanglement entropy as well as (in the ``probe" limit of
small tension) the high temperature free energy, confirming that the
two alternative definitions of the boundary entropy do indeed agree.

The second model we look at is the Janus solution
\cite{Bak:2003jk,Bak:2007jm}. In this example, one once more knows
the full bulk geometry and can calculate the entanglement entropy.
For Janus the dual DCFT is known and is of the interface type. We
can calculate the boundary entropy also in the limit of weak
coupling, where the calculation is tractable on the field theory
side. To leading order in the parameter controlling the jump across
the interface, we observe agreement between weak and strong
coupling.

Last, but not least, we look at defects with localized matter. These
systems often have a dual description in terms of a probe brane
embedded in the AdS$_3$ space. In these scenarios we don't have
access to the entanglement entropy without controlling the backreaction.
However, we can calculate the boundary entropy via the high
temperature free energy, as was already pointed out in
\cite{Yamaguchi:2002pa}. The dual field theory is once more well
understood and we can compare weak and strong coupling answers.

\subsection{Defect in a Toy Model}

Our first example of a 3-dimensional geometry with a dual
description in terms of a DCFT arises as a solution to the RS
\cite{Randall:1999vf} action of a 2d brane with tension $\lambda$ \be S =\frac{1}{16 \pi G_N^{(3)}} \int
d^3 x \sqrt{-g} (R +\frac{2}{R_{AdS}^2}) - \lambda \int d^2 x
\sqrt{-g_I}, \ee where $g_I$ is the determinant of the induced metric
on the 2d slice spanned by the brane. Without the brane the solution
to this system would be AdS$_3$ with curvature radius $R_{AdS}$. For
branes with a tension $\lambda$ less than a critical value $
\lambda_* = \frac{1}{4 \pi G_N^{(3)} R_{AdS}}$ one can find
solutions which have a brane with an AdS$_2$ geometry and hence
precisely preserve the symmetries expected from a dual DCFT. No
embedding of the system in this simple form into string theory is
known. Assuming that it makes
sense as a quantum theory, the observables in this theory have the
interpretation of correlation functions in some DCFT
\cite{Karch:2000ct}. For this toy model we do not have an
alternative definition of the DCFT.

\subsubsection{The Background Solution}
To construct the solution, consider the $d+1$ dimensional asymptotic
AdS background \be \label{adsads} ds^2= R_{AdS}^2(dy^2+e^{2A(y)}
(ds_{AdS_{d}})^2). \ee The pure AdS corresponds to $e^{A(y)}=\cosh
y$.

We are interested in the dual of a 2d CFT so we set $d=2$. Then we
can write $(ds_{AdS_{d=2}})^2=-\cosh ^2 r dt^2+dr^2$. In the
ordinary global coordinate we can rewrite as follows \be
\label{globalads} ds^2=R_{AdS}^2(-\cosh^2 \rho
dt^2+d\rho^2+\sinh^2\rho d \theta ^2). \ee For pure AdS$_3$, the coordinates are
related to each other via \be \cosh y \cosh r=\cosh\rho,\ \ \ \
\sinh y=\sinh\rho\sin\theta. \ee Using these global coordinates the
geometry on which the dual CFT lives is actually a circle and not
just a line. There are two defects at $\theta=0$ and
$\theta=\pi$, corresponding to the boundary points at fixed $y$ but
infinite $r$. In the presence of a codimension one defect with
tension $\lambda$, the equation of motion becomes
\be -1+(A')^2+A''=8\pi G^{(3)}_N R_{AdS} \lambda \delta(y),
\ee where we assumed that the brane is situated at $y=0$. This can
be solved by \be e^{A(y)}=\cosh(|y|-y_*). \ee The constant $y_*$ is
defined by \be \tanh y_*=4\pi G^{(3)}_N R_{AdS} \lambda. \ee

The spacetime with the backreaction due to the brane becomes two
copies of the partial AdS spacetime defined by $-y_*< y <\infty$ in
the AdS sliced coordinates (\ref{adsads}).

\subsubsection{Boundary Entropy from Entanglement Entropy}

As mentioned in the beginning of this section, the holographic
recipe for calculating entanglement entropies is to find at a fixed
time $t$ the minimal area surface in the bulk that ends on the
boundary of the region whose entropy we want to calculate. In the
case of a 3-dimensional bulk spacetime this minimal spatial
area is simply a geodesic. If we concentrate on the largest
region in the field theory that is symmetric around the defect we
are looking for a geodesic that connects the boundary points $\theta
= - \pi/2$ and $\theta=\pi/2$.
That is, in the coordinate system of eq. (\ref{adsads}) we
want to connect the point $r=0$ at $y = + \infty$ with the point
$r=0$ at $y=-\infty$. By symmetry it is easy to see that the
geodesic is $r=0$. We will return to this in more detail
later when we look at asymmetric regions.

For this longest geodesic we can easily calculate the extra length
$\Delta L$ induced by the defect brane as follows \be \Delta
L=2R_{AdS}y_*. \ee

Thus the extra contribution to the entanglement entropy becomes \be
\Delta S_A=\f{R_{AdS}y_*}{2G^{(3)}_N}. \ee In the probe limit,
the brane tension is very small  ($y_*\ll1$) and we can
approximately obtain \be \label{rsee} \Delta S_A=2\pi R_{AdS}^2
\lambda. \ee $\log g=\Delta S_A$ can directly be identified as the
boundary entropy of the dual DCFT.

\subsubsection{Boundary Entropy from Free Energy}
\label{btzrs}

Without the brane, turning on a finite temperature corresponds to
replacing the AdS$_3$ solution in the bulk with a BTZ black hole,
\be \label{BTZmetric} ds^2 = - h(r_{BTZ}) dt^2 +
\frac{dr_{BTZ}^2}{h(r_{BTZ})} + r_{BTZ}^2 d \theta^2 \ee with
$h(r_{BTZ}) = r_{BTZ}^2 - \mu +1$. For simplicity we switched to
units in which the curvature radius $R_{AdS}=1$.  For $\mu=0$ this
is simply global AdS and reduces to eq. (\ref{globalads}) by a
change of coordinates $\sinh(\rho) = r_{BTZ}$. The BTZ black hole
has a horizon at $r_H$ such that $r_H^2 = \mu-1$. The temperature
of the black hole is given by $T=\frac{h'(r_H)}{4 \pi} =
\frac{r_H}{2 \pi}$.

In order to study the free energy of the DCFT at finite temperature
we need to find the generalization of the BTZ black hole with the
backreaction of the brane included. This is a difficult problem and
no solutions are known. However, in the small tension limit the
change of the geometry due to the brane can be neglected. As a power
series expansion in the tension of the brane, the leading
contribution comes from the on-shell action of the brane probe which
minimizes its worldvolume in the fixed BTZ black hole background
geometry. This is in complete analogy to the calculation that allows
one to calculate order $N_f N_c$ corrections to the order $N_c^2$
free energies in a theory with a large number of colors $N_c$ and a
finite number of flavors $N_f$ using probe branes
\cite{Karch:2002sh}. This technique has been first used for a free
energy calculation in \cite{Babington:2003vm} and has been confirmed
by many calculations since.

The action describing the embedding of the brane is
proportional to the worldvolume of the brane, \be S_{probe} = -
\lambda \int d^2 x \sqrt{-g_I}. \ee A simple embedding is given by
the union of $\theta=0$ and $\theta = \pi$. This is the finite
temperature generalization of the probe stretching straight across
the AdS$_3$ space along the central $y=0$ slice in the coordinate
system of eq. (\ref{adsads}). It is the minimal action configuration
which satisfies the boundary conditions that the probe ends on the
defects\footnote{Alternatively we can work in the analog of Poincare
Patch coordinates where we drop the 1 from $h(r)$ and think of
$\theta$ as living on the real line as opposed to on a circle. In
this case we describe a single defect at $\theta=0$.}, which are
located at $\theta=0$ and $\theta=\pi$. The induced metric on this
brane is $ds^2 = - h dt^2 + \frac{dr_{BTZ}^2}{h}$ and so the
determinant of the induced metric is 1. Wick rotating to
Euclidean signature and regulating the on-shell action by simply
subtracting the zero temperature answer we get for the free energy
associated with a single defect  \be F = -T S_{on-shell} =
 \lambda  \lim_{r_c \rightarrow \infty} \left (
\int_{r_H}^{r_c}dr - \int_0^{r_c} dr \right ) = -\lambda r_H = -2
\pi T \lambda. \ee The entropy now can be calculated using the
standard relation \be \label{rsfe} S = - \frac{\partial F}{\partial
T} = 2 \pi \lambda.\ee Restoring the curvature radius $R_{AdS}$,
this is in perfect agreement with the answer for the boundary
entropy we got from the entanglement entropy, eq. (\ref{rsee}).

\subsection{Boundary Entropy and Janus Solution}

It is well-known that the near horizon limit of the D1-D5 system is
type IIB string theory on AdS$_3\times S^3\times T^4$. We assume
that there are $Q_1$ D1-branes and $Q_5$ D5-branes in this system.
The AdS$_3$/CFT$_2$ correspondence claims that the string theory in
this background is dual to the $(4,4)$ superconformal sigma model
whose target space is the symmetric product
$(T^4)^{Q_1Q_5}/S_{Q_1Q_5}$. We would like to deform this CFT so
that it includes a conformal defect. In particular, we are
interested in an interface which separates two regions with $T^4$ of
different radii. We assume that the radius of $T^4$ changes from
$R_+$ to $R_-$.

Recently, a 3 dimensional gravity background has been constructed  \cite{Bak:2007jm} that is a
particular example of the Janus solutions.
The supergravity metric in the Einstein frame is
\be ds^2_{IIB}=e^{\f{\phi}{2}}(ds^2_{(3)}+d\Omega_3^2)
+e^{-\f{\phi}{2}}ds^2_{T^4}.\label{sugrab} \ee

The $(2+1)$ dimensional part $ds^2_{(3)}$ is described by the
Einstein-Hilbert action plus a scalar field $\phi$ (i.e. in Einstein
frame). This is because $\s{-g^{(10)}}R^{(10)}=\s{-g^{(3)}}R^{(3)}$.
For the Janus solution, the 3D metric is explicitly given by \be
ds^2_{(3)}=R^2_{AdS}(dy^2+f(y)ds^2_{AdS_2}),\label{adsja} \ee where
the function $f(y)$ is found to be \be
f(y)=\f{1}{2}(1+\s{1-2\gamma^2}\cosh 2y). \ee Also, the two
asymptotic values of the dilaton $\phi_{\pm}\equiv\phi(\pm \infty)$
are found to be \be \phi_\pm=\phi_0
\pm\f{1}{2\s{2}}\log\left(\f{1+\s{2}\gamma}{1-\s{2}\gamma}\right).
\ee
For $\gamma=0$ the dilaton is constant, $\phi=\phi_0$, and the
metric reduces to pure AdS in the coordinate system of eq.
(\ref{adsads}).

The geodesic distance $L$ is needed in order to compute the
holographic entanglement entropy.  To obtain it, we have to be careful about the
regularization of the UV divergence. This can be done by expressing
the asymptotically AdS metric always in the form $ds^2_{(3)}\simeq
R^2_{AdS}\f{dz^2+dx^2-dt^2}{z^2}.$ Then the UV cutoff is always
given by $z=\ep$. In our case of (\ref{adsja}) we obtain \be
\ep=e^{-y_{\infty}}\f{2}{(1-2\gamma^2)^\f{1}{4}}. \ee In this way, we can
find the additional contribution to the entanglement entropy when we put a
non-zero value of the Janus deformation $\gamma$ to be \be
\Delta L=L-L_{\infty}=2R_{AdS}(y_{\infty}(\gamma)-y_{\infty}(0))
=R_{AdS}\log\f{1}{\s{1-2\gamma^2}}. \ee

The radius and the 3D Newton constant expressed in terms of the dual
2d CFT quantities are given by \be \label{d1d5map}
 R_{AdS}=\s{g_6}(Q_1Q_5)^{1/4}l_s,\ \ \ \
4G^{(3)}_N=\s{g_6}(Q_1Q_5)^{-3/4}l_s, \ \ \
g_6=g_s\s{\f{Q_5}{Q_1}}.\ee

Thus, in the end we obtain the shift of the entanglement entropy as
follows: \be \Delta S_A=\f{\Delta
L}{4G^{(3)}_N}=\f{Q_1Q_5}{2}\log\f{1}{1-2\gamma^2}
=Q_1Q_5(\gamma^2+\gamma^4+\ddd).\label{resultads} \ee We can claim
that this finite part which appears in the Janus background actually
corresponds to the boundary entropy (or the logarithm of the
$g$-function) by applying the relation (\ref{entg}).

Now we want to perform the direct computation of the boundary
entropy from the CFT side in order to compare with the above
result.
To treat the defect CFT we need the doubling trick discussed in
\cite{Bachas:2001vj}. Consider again a single compactified scalar
$\phi$ in the presence of the interface where the radius of the
scalar jumps from $R_+$ to $R_-$. This theory is equivalent to a
BCFT with two scalar fields whose radii are $R_+$ and $R_-$. The
boundary condition is the Neumann-Dirichlet type (i.e. there is a `D1-brane'
which wraps the diagonal $S^1$ in $T^2$) as we will review in
appendix A. Since the $g$-function is proportional to the tension of
the D-brane and is T-dual invariant, we obtain (see
\cite{Elitzur:1998va}) the following results for a single boson
compactified with the radius $\ti{R}$ \be
g_{N}=\s{\f{\ti{R}}{\s{2\al}}},\ \ \ \ \
g_{D}=\s{\f{\s{\al}}{\s{2}\ti{R}}}, \ee for Neumann and Dirichlet
boundary conditions, respectively. For the Dp-brane wrapped on $T^p$ with the
B-field (i.e. the gauge flux) we obtain \be g=2^{-p/4}\cdot\det
(G-BG^{-1}B), \label{gdp} \ee where we assume that all torus
coordinates have the periodicity $x_{i}\sim x_i+2\pi\s{\al}$.

Our system is described by a D1-brane stretching in the diagonal
direction of $T^2$. This is T-dual to a D2-brane with a gauge flux
$B_{12}=1$, which corresponds to a single D0 charge.
Plugging in $g_{11}=\f{1}{R_1^2}$ and $g_{22}=R_2^2$, the formula
(\ref{gdp}) leads to \be
g=\f{1}{\s{2}}\s{\f{R_+}{R_-}+\f{R_-}{R_+}}. \ee Indeed we can
confirm $g=1$ at $R_+=R_-$, which corresponds to the absence of the
defect. Thus we
get \footnote{The boundary entropy of this DCFT very
recently has also been studied in \cite{Bachas:2007td}.
Interestingly, the authors find that such an interface,
where the radius of a compact scalar jumps by a
finite amount, can increase the entropy by splitting into 2 defects
with smaller jumps. This process repeats and
ultimately one should obtain infinitely many
defects with infinitessimally small jumps.
In \cite{Bachas:2007td} this property is identified as an
instability in the sense of the renormalization group flow. The
CFT has relevant operators that drive the RG flow away from the
fixed point with a single defect. This only turns
into a dynamical instability if we promote
the radius of the scalar into a dynamical field. We thus should not expect to see
this as an instability in the spectrum of normalizable fluctuations
around the Janus geometry with fixed asymptotic behavior,
consistent with the positive energy theorem
proven for Janus-type solutions in \cite{Freedman:2003ax}.
  }
 \be \Delta S_{bdy}=\log g=
\log\f{\s{\f{R_+}{R_-}+\f{R_-}{R_+}}}{\s{2}}. \ee

Then we need to estimate the value of $\f{R_+}{R_-}$ dual to the
Janus solution. First, we notice that the warp factor of the $T^4$
part becomes the constant $1$ in the {\it string} frame because
$G^{Einstein}_{\mu\nu}=e^{-\f{1}{2}\phi}G^{string}_{\mu\nu}$. Thus
the kinetic term of the $(T^4)^{Q_1Q_5}/S_{Q_1Q_5}$ sigma model
should be proportional to $\f{1}{g_s}=e^{-\phi}$. Explicitly, this
term goes like $\sim e^{-\phi}\int dz^2 G^{string}_{\mu\nu}\de X^\mu
\bar{\de} X^\nu$. Thus the radius is proportional to $e^{-\phi/2}$.
The ratio $R_+/R_-$ in the Janus solution becomes \be
\f{R_+}{R_-}=\left(\f{1+\s{2}\gamma}{1-\s{2}\gamma}\right)^{\f{1}{2\s{2}}}.\label{ratior}
\ee We can estimate the boundary entropy as follows \be
S_{bdy}=4Q_1Q_5\log\f{1}{\s{2}}
\s{\left(\f{1+\s{2}\gamma}{1-\s{2}\gamma}\right)^{\f{1}{2\s{2}}}+
\left(\f{1-\s{2}\gamma}{1+\s{2}\gamma}\right)^{\f{1}{2\s{2}}}}=
Q_1Q_5(\gamma^2+\f{7}{6}\gamma^4+\ddd). \label{resultcft}\ee As
expected, the boundary entropy in the free theory can also be
calculated via the free energy yielding identical results. We
present that calculation in appendix \ref{appendix2point}.

\begin{figure}[htbp]
\begin{center}
  \hspace*{0.5cm}
  \includegraphics[height=4cm]{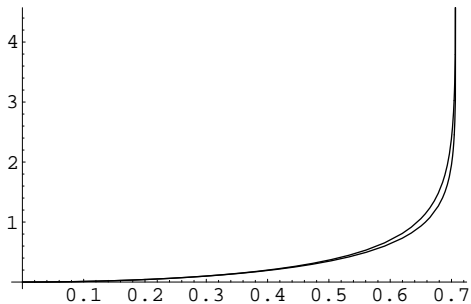}
  \caption{The plot of the boundary entropy from both the AdS and free CFT calculation.
We plotted the values of $\f{S_{bdy}}{Q_1Q_5}=\f{\log g}{Q_1Q_5}$ as
a function of $\gamma$. Notice that $\gamma$ can take the values
$0\leq \gamma \leq \f{1}{\s{2}}.$ The upper and lower curve
correspond to the free field CFT result (\ref{resultcft}) and AdS
result (\ref{resultads}), respectively. They almost coincide with
each other, but there is a small deviation. }\label{comp}
\end{center}
\end{figure}

Thus the leading term ($\sim \gamma^2$) from AdS (\ref{resultads})
agrees with the one from CFT (\ref{resultcft}). Thinking of the
Janus field theory in the framework of conformal perturbation
theory, as in \cite{Clark:2004sb}, this agreement hints at a
non-renormalization of some correlation functions of the Lagrangian.  
The relevant correlation functions are those of the Lagrangian with the twist fields that produce the $n-$sheeted
Riemann surface, corresponding to $Z_n$ in (\ref{formulaz}); refer
to \cite{Calabrese:2004eu,Ryu:2006ef} for general discussion. Also,
as shown in fig.\ref{comp}, the deviations of (\ref{resultads}) from (\ref{resultcft}) are
very small for any value of $\gamma$. We may notice that the
boundary entropy in the free field theory is always larger than that
in AdS (i.e. at strong coupling), which is natural.

\subsection{Probe Computations for D1 D5 System}

Last, but not least, we want to study another set of DCFTs. A whole
class of DCFTs can be realized via probe\footnote{ Probe brane here
refers to the limit where the backreaction of the brane on the
geometry is negligible, so that the brane simply minimizes its
worldvolume action in a given fixed background. In the field theory
this corresponds to the quenched approximation which is justified by
a large number of colors limit.} D-branes in known AdS backgrounds
\cite{Karch:2000gx}. For AdS$_3$ such probe branes
were first discussed in \cite{Bachas:2000fr}.
In the dual field theory, the probe brane
corresponds to adding a finite number of localized matter fields
into a CFT with a large number of degrees of freedom, for example
$N_f$ fundamental hypermultiplets into a large $N_c$ gauge theory.
Focusing on 2d field theories, we can start with the AdS$_3\times
S^3\times T^4$ spacetime considered in the last subsection and add a
probe F1-string on AdS$_2$ or a probe D3 brane
on AdS$_2 \times S^2$ \cite{Raeymaekers:2006np, Couchoud:2003jw}. Without solving for the backreaction of these
branes we cannot extract the entanglement entropy. It is however
straightforward to obtain the free energy at high temperatures
directly from the probe action, just as we did in section \ref{btzrs}
for the RS brane in the small tension limit.

Consider $M$ F1-strings or $M$ D3 branes in the BTZ black hole
background eq. (\ref{BTZmetric}), since these are the potential
supersymmetric probes. These branes have an action which is given by
their area just as in the RS toy model. In addition there are also
couplings to the background form fields, in particular in the WZ
term in the D3 action. We must consider how these terms contribute to the action. In the
string theory setup the background is supported by a 3-form RR flux
$H$. A convenient gauge choice for the RR 2-form is to take it to be
of the form $B_{RR} = B(r) dt \wedge d\theta$. The probe embedding
we are looking for is the same $\theta=0$ one we discussed in the RS
toy model. Since the pullback of $B_{RR}$ for this
embedding is zero it does not contribute to the on-shell action and,
as before, we get the free energy (and hence the boundary entropy
associated with the defect) simply from the volume of the brane. The
result, as in eq. (\ref{rsfe}), is that for a brane of tension
$\lambda$ the boundary entropy of the defect dual to the probe brane
is \be S = 2 \pi R_{AdS}^2 \lambda. \ee For the F1 and D3 branes all
we need is to plug in the relevant values of the tension
$\lambda_{F1,3}$ using eq. (\ref{d1d5map}). The general formula is
that the tension of a Dp brane is $\lambda_p = \frac{1}{(2 \pi)^p}
\frac{1}{gs} \frac{1}{l_s^{p+1}} $ and $\lambda_{F1}=\frac{1}{2\pi
l_s^2}$ for the fundamental string. With this we get for a single
probe brane
\begin{eqnarray}
\nonumber  \lambda_{F1} = \frac{1}{2 \pi l_s^2} = \frac{g_s Q_5}{2
\pi R_{AdS}^2} \,\,\, &\Rightarrow \,\,\,& S_{F1} = g_s Q_5, \\
 R_{AdS}^2 \lambda_3 = \frac{R_{AdS}^2}{(2 \pi)^3 g_s l_s^4} =
\frac{g_s Q_5^2}{(2 \pi)^3 R_{AdS}^2} \,\,\, &\Rightarrow \, \, \,&
S_3 = \frac{g_s Q_5^2}{(2 \pi)^2}.
\end{eqnarray}
For $M$ probe branes we get $M$ times these expressions. A
boundary entropy scaling as $M Q_5$ is expected for the D3 brane
from the weak coupling consideration. This gets enhanced by a power
of the 't Hooft couplings $g_s Q_5$. Such a strong coupling
enhancement of the free energy has been seen in other probe systems
before, such as the D7 probe that adds flavor to ${\cal N} =4$
super-Yang-Mills, where the free energy scales as $\lambda N_f N_c$
instead of the naive $N_f N_c$ (see e.g.
\cite{Mateos:2006nu,Albash:2006ew,Karch:2006bv}). For the F1 string
we see a very similar effect. This determination of the boundary
entropy from the free energy contribution due to a probe brane can
easily be generalized to higher dimensional systems. What is unclear
to us at the moment is whether, as in 2d, in these higher
dimensional examples an equivalent definition of the boundary
entropy can also be given via the entanglement entropy. We hope to
return to this issue in the future.

\subsection{Size and Shape (In)dependence}

Given our understanding of the meaning of the boundary entropy, we
would expect that the contribution to the boundary entropy of a
defect should be independent of the size of the subsystem enclosing
the defect as long as the defect is in the center of the interval.
For a defect that is off-center the folding trick can not be used to
reduce the entanglement entropy calculation in the DCFT to the well
known case of a BCFT, as we pointed out in section \ref{befromee}.
It appears that in a DCFT the entanglement entropy of such an
asymmetrically shaped region depends explicitly on the microscopic
details and not just on the two universal numbers $c$ and $g$. In
this subsection we want to reanalyze this issue in the context of
the holographic calculation in the Janus framework.

We calculate the entanglement entropy of a spatial interval of
length $l$ containing the defect on the field theory side, but
potentially off-center. From the three-dimensional point of view, we
must find the geodesic length of a spacelike segment $r(y)$ in the
metric (\ref{adsads}) with one endpoint at $y = -\infty$, $r = r_0$
and one endpoint at $y = \infty$, $r = r_0 + \Delta r$. In line with
what we said above, we expect the boundary entropy to be independent
of the length $r_0$ but to depend on the asymmetry of the interval
about the defect parameterized by $\Delta r$.

The geodesic action is independent of $r$ and leads to the
conservation equation
\begin{equation}
\frac{f(y) r'}{\sqrt{f(y) r'^2 +1}} = \alpha
\end{equation}
where $\alpha$ is some constant that sets the asymmetry of the
interval that is nontrivially related to $\Delta r$ (this can be seen by integrating $r'$ from $y = -\infty$ to
$y = \infty$).

The fact that the geodesic length depends only on $r'$ tells us
immediately that the boundary entropy is independent of $r_0$ as
expected. In order to establish the dependence on $\alpha$, we must
calculate the geodesic length in the Janus system for some nonzero
$\alpha$ and subtract from it the geodesic length in the pure AdS
system, being careful that in both calculations the boundary
interval has the same length. It is very easy to see that in this
case the difference in entanglement entropies between Janus and pure
AdS gets a contribution from the detailed shape of the warpfactor
around the center of AdS (that is, around $y=0$).

\section{Conclusion}
\label{conclusion}
In this paper we have calculated the boundary entropy in several
strongly coupled 2d defect conformal field theories which have a
holographic dual. We confirmed that the definition of the boundary
entropy in terms of the entanglement entropy gives identical answers
to the definition in terms of a free energy at large temperature.
Perhaps most interestingly, we found that this equivalence only holds
in the case that one calculates the entanglement entropy for an
interval that has the defect at the center, so that the DCFT can be
mapped to a BCFT via the folding trick and the entanglement entropy
is completely specified by two universal numbers, the boundary
entropy and the central charge. In a DCFT, the entanglement entropy
of an asymmetric interval captures detailed information about the
microscopic details of the theory. In particular, from the knowledge
of the entanglement entropy for arbitrarily shaped intervals one can
reconstruct the length of all geodesics in the bulk and hence
presumably the bulk metric.

Our methods employed in the bulk can readily be generalized to
higher dimensions. In this case it is not clear if there is a
similar universal definition of a boundary entropy as in 2d, though
we may speculate that a coefficient of subleading divergent parts in
the entanglement entropy will be a counterpart of the boundary
entropy. However, it should still be interesting to calculate free
energies and entanglement entropies associated with defects in
strongly coupled theories in more than 2 dimensions.

\vskip2mm

\noindent {\bf Acknowledgments}

AK would like to thank the Yukawa Institute for Theoretical Physics
in Kyoto for their hospitality while this work was initiated. The
work of AK and ET was supported in part by the U.S. Department of
Energy under Grant No. DE-FG02-96ER40956. The work of TT is
supported in part by JSPS Grant-in-Aid for Scientific Research
No.18840027 and by JSPS Grant-in-Aid for Creative Scientific
Research No. 19GS0219.

\appendix

\section{Janus Entropy at Weak Coupling from Free Energy}
\label{appendix2point} \setcounter{equation}{0}

\subsection{Two Point Function in the Presence of the Defect}
We consider the interface CFT defined by a $D$ dimensional free
scalar field $\phi$ whose radius jumps from $R_+$ to $R_-$ at $y=0$.
This is defined by the following action (in Euclidean space)
\begin{eqnarray}
S = \frac{R_{-}^2}{2}\int_{y<0} d^{D-1}xdy
\partial_{\mu}\phi_-\partial_{\mu}\phi_- +\frac{R_{+}^2}{2}\int_{y>0}
d^{D-1}xdy \partial_{\mu}\phi_+\partial_{\mu}\phi_+
\end{eqnarray}
where $y$ is the direction which is perpendicular to the defect, $D$
is the total dimension and  $\mu $ runs over all spacetime directions except
$y$.  We define \be c_{\pm}=\f{1}{R^2_{\pm}}.\ee Imposing the
on-shell condition, variation of the action with respect to $\delta \phi_{\pm}$ leads to \be \delta S=\left[\f{1}{c_+}\de_y\phi_+\delta\phi_+
-\f{1}{c_-}\de_y\phi_-\delta\phi_-\right] |_{y=0}.\ee Since we
require the boundary condition \be
\phi_+(x_\mu,y=0)=\phi_-(x_\mu,y=0), \label{bodc} \ee we have
$\delta \phi_+=\delta \phi_-$ at $y=0$. Thus the principle of least
action leads to \be c_- \de_y\phi_+(x_\mu,y=0) = c_+ \de_y
\phi_-(x_\mu,y=0).\label{bodcc} \ee These two conditions are enough
to determine the propagators \ba \la
\phi_{+}(x_1,y_1)\phi_{+}(x_2,y_2)\lb&=&
\f{c_+}{\left((y_1-y_2)^2+(x_1-x_2)^2\right)^{\f{D-2}{2}}}
+\f{c_+a_+}{\left((y_1+y_2)^2+(x_1-x_2)^2\right)^{\f{D-2}{2}}},\no
\la\phi_{-}(x_1,y_1)\phi_{+}(x_2,y_2)\lb&=&
\f{c_+(1+a_+)}{\left((y_1-y_2)^2+(x_1-x_2)^2\right)^{\f{D-2}{2}}},\\
\la\phi_{+}(x_1,y_1)\phi_{-}(x_2,y_2)\lb&=&
\f{c_-(1+a_-)}{\left((y_1-y_2)^2+(x_1-x_2)^2\right)^{\f{D-2}{2}}},\no
\la\phi_{-}(x_1,y_1)\phi_{-}(x_2,y_2)\lb&=&
\f{c_-}{\left((y_1-y_2)^2+(x_1-x_2)^2\right)^{\f{D-2}{2}}}
+\f{c_-a_-}{\left((y_1+y_2)^2+(x_1-x_2)^2\right)^{\f{D-2}{2}}},
\label{propa} \nonumber \ea where we defined \be a_{+}=-a_{-}=\f{c_-
-c_+}{c_- + c_+}.\ee

It is clear from the above formulation that we can equivalently
treat the system such as two fields $\phi_{+}$ and $\phi_{-}$
that live in the same half space defined by $y\geq 0$. This is called the
doubling (or folding) trick and it is rather common, especially in
two dimensional CFTs. From this perspective, the constraints (\ref{bodc})
and (\ref{bodcc}) are regarded as the Neumann-Dirichlet boundary
condition at the open boundary $y=0$.

\subsection{The Boundary Entropy in the Presence of the Defect}
Now we concentrate on the $D=2$ case. In order to compute the
boundary entropy, we need to evaluate the partition function. We
employ the normalized field $\vp_{\pm}=R_{\pm}\phi_{\pm}$ such that
the action looks like \ba S&=&\int_{y>0} dtdy [\de_{\mu}\vp_+
\de_{\mu}\vp_+]+\int_{y<0} dtdy [\de_{\mu}\vp_- \de_{\mu}\vp_-]\no
&=& -\int_{y>0} dtdy [\vp_+ \de_{\mu}\de_{\mu}\vp_+]-\int_{y<0} dtdy
[\vp_- \de_{\mu}\de_{\mu}\vp_-].\ea The boundary condition now
becomes \be R_-\vp_+=R_+\vp_-,\ \ \ R_+\de_{y}\vp_+=R_-\de_{y}\vp_-
\label{bfd} \ee at the interface $y=0$. Imposing (\ref{bfd}) and the
following normalization, \be \int^\infty_{0}
\vp^{(i)+}_p(y)\overline{\vp^{(j)+}_q(y)}
\f{dy}{2\pi}+\int^0_{-\infty}
\vp^{(i)-}_p(y)\overline{\vp^{(j)-}_q(y)} \f{dy}{2\pi}
=\delta(p-q)\delta_{ij}, \ee we obtain the orthogonal basis with the
momentum $p>0$  as follows\footnote{It is useful to note the step
function is expressed as follows \be \theta(y)=\f{1}{2\pi}\int
\f{dp}{ip+\ep}e^{ipy}, \ee which leads to \be \int^\infty_{0}
\f{dy}{2\pi}
e^{ipy}=\f{1}{-ip+\ep}=\f{1}{2}\delta(p)+i\f{p}{p^2+\ep^2}. \ee
Notice also \be \delta(p)=\f{1}{-ip+\ep}+\f{1}{ip+\ep}. \ee}  \ba
\vp^{(1)_+}_p(y)&=&\f{\nu-1}{\s{2(1+\nu^2)}}e^{ipy}+\f{\nu+1}{\s{2(1+\nu^2)}}e^{-ipy},\no
\vp^{(1)_-}_p(y)&=&\f{1-\nu}{\s{2(1+\nu^2)}}e^{ipy}+\f{\nu+1}{\s{2(1+\nu^2)}}e^{-ipy},
\label{transone} \ea and \ba
\vp^{(2)_+}_p(y)&=&\f{\nu+1}{\s{2(1+\nu^2)}}e^{ipy}+\f{\nu-1}{\s{2(1+\nu^2)}}e^{-ipy},\no
\vp^{(2)_-}_p(y)&=&\f{1+\nu}{\s{2(1+\nu^2)}}e^{ipy}+\f{1-\nu}{\s{2(1+\nu^2)}}e^{-ipy}.
\label{transtwo} \ea Here, we
defined $\nu=\f{R_+}{R_-}$ and the dependence on the time
$x^0=t$ has been suppressed.

When we expand the scalar field in terms of this basis \be
\vp=\sum_{p>0}\left(c^{(1)}_{p}\vp^{(1)}_{p}
+c^{(2)}_{p}\vp^{(2)}_{p}\right)+c_0\vp_0, \ee then the measure of
the path-integral is given by \be [D\vp]=
\left(\prod_{p>0}[dc^{(1)}_p][dc^{(2)}_p]\right)\cdot [dc_0]. \ee

On the other hand, in the case of the ordinary scalar field theory
without any interface, the normalized basis is given by $e^{ipx}$
(just setting $\nu=0$). The measure for this basis is denoted as \be
[D\vp^{0}]=\left(\prod_{p>0}[dc^{0(1)}_p][dc^{0(2)}_p]\right)\cdot
[dc^0_0]. \ee

It is easy to see $[dc^{0(i)}_p]=[dc^{(i)}_p]$ from (\ref{transone})
and (\ref{transtwo}). In this way we have found that the difference
between the partition function with and without the defect comes
from the $p=0$ contribution. Notice that the zero-mode $c^{(i)}_0$
spans the interval \be 0\leq c^{(i)}_0\leq \s{2}\pi \s{R_+^2+R_-^2}.
\ee This is because the scalar $\vp_{\pm}$ is compactified with the
radius $R_{\pm}$.

What we are interested in is the ratio \be
g=\f{Z_{interface}}{\s{Z_{R_+}Z_{R_-}}}, \ee where $Z_{R_+}$ is the
partition function with an infinitely long $x$ direction radius with
$R_+$. This clearly coincides with the ground state degeneracy $g$
discussed in this paper. In this ratio, the nonzero-modes, i.e.
$c^{(i)}_p$, cancel out completely. The zero-mode contribution in the
$x$ direction reads \be g=\f{ \s{R_+^2+R_-^2}}{\s{(\s{2}R_+)\cdot
(\s{2}R_-)}}=\s{\f{R_+^2+R_-^2}{2R_+R_-}}. \ee

\bibliography{entangle}
\bibliographystyle{JHEP}

\end{document}